\documentclass[twocolumn]{aastex61}

\usepackage{verbatim, graphicx,  ifthen, pbox}
\usepackage{eso-pic}
\usepackage{CJKutf8}
\usepackage{xcolor}
\usepackage{hyperref}
\usepackage{breakurl}
\usepackage{amsmath}
\usepackage{graphicx}
\usepackage{verbatim}
\usepackage{booktabs}
\usepackage[T1]{fontenc}
\usepackage{ae,aecompl}

\usepackage{newtxtext,newtxmath}

\begin{document}
\title{A General Origin for Multi-Planetary Systems With Significantly Misaligned USP Planets}


\author[0000-0001-9256-5508]{L.~Brefka}
\affiliation{Department of Astronomy, University of Michigan, Ann Arbor, MI 48109, USA}
\author[0000-0002-7733-4522]{J.~C.~Becker}
\affiliation{Division of Geological and Planetary Sciences, California Institute of Technology, Pasadena, CA 91125, USA}

\begin{abstract} 
	Ultra-short period (USP) planets are exoplanets which have orbital periods of less than one day and are unique because they orbit inside the nominal magnetic truncation gap of their host stars. In some cases, USP planets have also been observed to exhibit unique dynamical parameters such as significant misalignments in inclination angle with respect to nearby planets. In this paper, we explore how the geometry of a multi-planet system hosting a USP planet can be expected to evolve as a star ages. In particular, we explore the relationship between the {mutual} inclination of the USP planet and the quadrupole moment ($J_2$) of the host star. We use secular perturbation theory to predict the past evolution of the example TOI-125 system, and then confirm the validity of our results using long-term N-body simulations. Through investigating how the misalignment between the candidate USP planet and the three other short-period planets in the TOI-125 system arose, we intend to derive a better understanding of the population of systems with misaligned USP planets and how their observed parameters can be explained in the context of their dynamical histories. 
\end{abstract} 

\section{Introduction} \label{sec:intro}
The space-based transit missions Kepler, K2, and TESS have discovered thousands of planets and found a large number of new planetary orbital architectures. One of the most remarkable results coming from TESS in particular is a deeper understanding of the covariances in planet formation, as TESS’ bright targets enable follow-up efforts, which allow more complete measures of which types of planets are most likely to occur together within a single system. 
Previously, Kepler and K2 found that exoplanets frequently reside in multi-planet systems. Systems of multiple planets residing in a fairly narrow orbital plane and within a fairly short ($a<0.5$ AU) orbital radius have an occurrence rate of roughly $\sim$20-30\% \citep{Muirhead2015,Zhu2018}. 
These systems are observed to have greater degrees of intra-system regularity compared to the entire exoplanet sample \citep{Weiss2018, Millholland2017}.
Simultaneously, these tightly packed systems of inner planets (STIPs) can also host ultra-short-period planets \citep{Winn2018,Adams2020}. 

Ultra-short-period (USP) planets (defined as those with orbital periods of less than a day) occur around $\sim$0.5\% of G-dwarf stars \citep{SanchisOjeda2014}, and when they occur in STIPs, they often exist in apparently decoupled dynamical modes as compared to the rest of the system, sometimes in a coplanar configuration \citep[ex:][]{Swift2013, Becker2015, MacDonald2016} and sometimes in a misaligned configuration \citep[ex:][]{Rodriguez2018, Quinn2019}. 
Since the (sometimes near-resonant) geometry of STIPs indicates that they formed via disk migration \citep{Rein2012, Batygin2015, Deck2015}, the presence of an interior planet in a different dynamical mode {requires an explanation \citep{Pu2019, Petrovich2018, Millholland2020, Becker2021}.}
\citet{Dai2018} noted that USP planets' orbital inclinations have a wider range of values with respect to nearby, exterior planets than do similar but longer-period planets. The finding of \citet{Dai2018} suggests a mechanism producing these mutual inclinations that is orbital-radii-dependant.

Recently, work by \citet{Becker2020} studied the K2-266 system and suggested the misalignment between the STIP and its interior, misaligned USP planet could arise post-disk migration in two ways: via the presence of an exterior planetary companion at very particular orbital locations, or by dynamical evolution during the spin-down of the host star when some initial stellar obliquity is present \citep{Spalding2016, Li2020}. The latter explanation, if true, would create ubiquitous misalignments in systems containing USP planets and other nearby planets with slightly longer orbital periods, consistent with the correlation discussed in \citet{Dai2018}.

Already, TESS has provided several new examples of multi-planet systems that host USP planets, which give new tests of the explanations of the dynamics for systems with similar geometry that came from Kepler/K2 data. The TOI-125 system is a good example of this. TOI-125 has three confirmed planets and one additional planet candidate, and the innermost planet candidate is misaligned by about 14 degrees compared to the other planets \citep{Quinn2019, Nielsen2020}. For K2-266, TOI-125, and other yet-undiscovered systems with similar geometries, the inclination structure seen in the planetary system leaves a clue as to its past evolution. 

In this work, we consider the evolution of a multi-planet system with a misaligned USP in the presence of an evolving star, with a goal of describing how relative oscillations of planetary orbits change as the stellar non-spherical potential decreases with time. 
We use the measured parameters of the TOI-125 system as our test case, but the conclusions of this work apply more generally to systems with the TOI-125/K2-266 archetypal geometry as well.
First, in Section \ref{sec:secular}, we describe the TOI-125 problem using a secular model and make secular predictions for stellar $J_2$ values that denote transitions in the dynamics of the system. 
In Section \ref{sec:nbody}, we perform full N-body simulations to evaluate the continuous evolution of the TOI-125 planetary system under a changing stellar potential, and demonstrate the effect of stellar obliquity coupled with stellar spin-down. 
In Section \ref{sec:discuss}, we discuss the implications of our results on the exoplanet sample in general, and conclude in Section \ref{sec:conclude} with a summary of our results.

\section{{Semi-}Analytical Predictions}
\label{sec:secular}
\citet{Li2020} and \citet{Becker2020} suggest that the evolution of the stellar potential can alter the character of the planet-planet interactions in a planetary system containing a STIP with a USP planet. In particular, as the star ages and spins down, the quadrupole moment ($J_2$) which describes the oblateness of the star will decrease. 
The impact of the $J_2$ moment is to induce a nodal precession on nearby planets. 
As the star ages, this can result in planet-planet interactions becoming dominant over the stellar-induced precession for short-period planets. 
This will occur if the precession term (due to $J_2$) of the Hamiltonian is small in comparison to the planet-planet term for a particular planet.
We can use analytical methods to make predictions about at what values of stellar $J_{2}$ these transitions would have occurred in the TOI-125 system. 

We note that of the five candidate planets originally announced in \citet{Quinn2019},  only three have been confirmed \citep{Nielsen2020}.
\citet{Nielsen2020} finds an upper mass limit for the other two candidates, and finds that while TOI-125.05 is no longer a likely viable planet candidate. {At the time this paper was written,} TOI-125.04 remains a viable candidate with a consistent radius measurement and mass upper limit. 
In this work, we include in our analysis TOI-125 b, c, d and .04 and use the orbital parameters from \citet{Quinn2019} and \citet{Nielsen2020}. The dynamics discussed in this work is applicable to any system with a similar geometry.
We use TOI-125 as an illustrative example of this geometric archetype. {We note that even if TOI-125.04 is proven to be a false positive, then the dynamics described in this paper can easily be applied to other systems with similar geometries such as K2-266 or any future systems discovered to have similar properties.}

\begin{figure}[h!]
    \centering
    \includegraphics[width=3.4in]{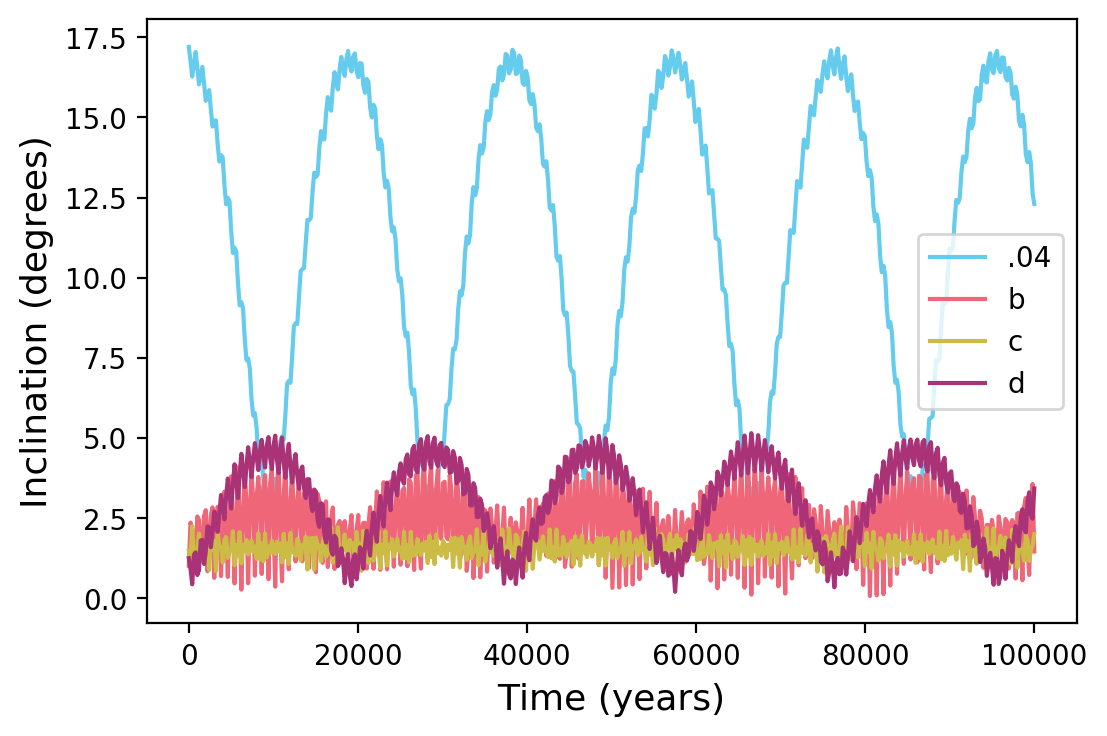}
    \caption{Plot of the secular inclination evolution over $10^{5}$ years using the current-day measured orbital elements for the TOI-125 planets for a singular $J_2$ value of $3.3 \cdot 10^{-6}$. This timescale was chosen in order to allow for visualization of several secular cycles, and the anti-correlation of the oscillations illustrates the planet-planet interactions that can be seen in the system. }
    \label{fig:1e-04SecInc}
\end{figure}
\subsection{The Secular Theory}
In the absence of changes in planetary semi-major axis or orbital resonance, the dynamics of an exoplanet system can be well-approximated using Laplace-Lagrange secular theory \citep{MD99}.
In order to represent the dynamical interactions between the planets of TOI-125, and to understand the evolution of their inclinations over time, we apply {the secular theory} to the disturbing function. Essentially, this entails treating the planets as their time-averaged potentials, and evolving a system of rings representing the orbits. {The approximation holds for small planetary eccentricities and inclinations ($<$30 degrees).}  We can write the disturbing function to second order in terms of planetary inclinations as 
    	\begin{equation}
	\begin{split}
	     \mathcal{R}_{j}^{(sec)}= n_ja_j^2[
	     \frac{1}{2}B_{jj}I_j^2 
	    +\sum_{k=1}^N B_{jk}I_jI_k cos(\Omega_j-\Omega_k)], \\
	\end{split}   
	\end{equation} 
    where $n$ is the planetary mean anomaly, $j$ the planet index whose evolution is being computed, $k$ the index of the perturbing planet, $I$ is the inclination, $\Omega$ is the longitude of ascending node, and $a$ is the semi-major axis. {Note that the full second-order secular theory also can compute the evolution of planetary eccentricities, but we neglect those terms in our analysis since we are concerned with planetary inclinations. The full disturbing function can be found in \citet{MD99}.} The coefficients $B_{jj}$, and $B_{jk}$ can then be written as 
    \begin{equation}
    \begin{split}
        B_{jj}=-n_j[\frac{3}{2}J_2(\frac{R_c}{a_j})^2-\frac{27}{8}J_2^2(\frac{R_c}{a_j})^4 \\
        +\frac{1}{4}\frac{m_k}{m_c+m_j}\alpha_{jk}\overline{\alpha}_{jk}b_{3/2}^{(1)}(\alpha_{jk})], \\
    \end{split}
    \label{eq:intrinsicfreq}
    \end{equation}
    \begin{equation}
         B_{jk}=+n_j\frac{1}{4}\frac{m_k}{m_c+m_j}\alpha_{jk}\overline{\alpha}_{jk}b_{3/2}^{(1)}(\alpha_{jk}), \\
    \end{equation}
    where $J_2$ describes the oblateness of the star, $R_c$ is the radius of the star, $\alpha_{jk}$ is {the ratio of the inner planet's semi-major axis to the outer planet's semi-major axis ($\frac{a_j}{a_k}$ when $a_j <a_k$ and $\frac{a_k}{a_j}$ otherwise) and $\overline{\alpha}_{jk} = \alpha_{jk}$ except when $a_j <a_k$, in which case $\overline{\alpha}_{jk} = 1$. }
    The parameter $J_2$ can also be written in terms of the star's physical parameters as follows \citep[][]{Sterne1939, Ward1976}: 
\begin{equation}
J_2 = \frac{k_2}{3} \left(\frac{\Omega_*}{\Omega_{*,b}}\right)^2 
\end{equation}
{where $\Omega_{*}$ is the stellar rotational frequency, $\Omega_{*,b}$ is that frequency at break-up}, and $k_2$ is the love number of the star. In general, as the spin of a star decreases, its quadrupole moment $J_2$ will also decrease. 
    The value $b_{3/2}^{(1)}$ denotes a Laplace coefficient, defined as 
    \begin{equation}
    \begin{split}
        b_{3/2}^{(n)}(\alpha)=\frac{1}{\pi}\int_{0}^{2\pi}\frac{cos(n\psi)d\psi}{(1-2\alpha cos(\psi)+\alpha^2)^\frac{3}{2}},
    \end{split}
    \end{equation}
The $B$ coefficients above are frequencies which describe the interactions between the pairs of planets in the system (and themselves), and are parts of their respective matrix ${B}$, which describes the inclination evolution of the planets. These matrices can be solved for their eigenfrequencies and eigenvectors in order to determine this evolution. For ${B}$ {they are respectively labeled} $f_k$ and $\Bar{e_k}$. We then define {variables} in order to describe the time evolution of the system in terms of inclination and {longitude of the ascending node}:
\begin{equation}
    p_j= I_j sin~\Omega_j,~ q_j = I_j cos~\Omega_j
\end{equation}
The time dependent solutions for these equations then take the form
\begin{equation}
    p_j(t) = \sum_{k=1}^N I_{jk} sin(f_k t +\gamma_k)
\end{equation}
\begin{equation}
    q_j(t) = \sum_{k=1}^N I_{jk} cos(f_k t +\gamma_k)
\end{equation}
Where the phases ($\gamma_k$) are determined by the initial conditions. Our true inclination values can be determined through the equation 
\begin{equation}
    I_{jk} = \mathcal{I}_{jk}T_{jk} 
\end{equation}
The constant $T_{jk}$ being determined once more through the boundary conditions, which are in most cases the current-day best-fit values of the planet parameters found by an analysis of the transit data obtained by TESS \citep{Quinn2019} and follow-up radial velocity constraints from \citet{Nielsen2020}, which are reproduced in Table \ref{tab:ParamsTable} of this work. To reduce the degeneracy in our solutions, we consider only the best-fit solution from the observational fit and consider it representative of the observational posteriors. 

After solving {Equations 1-9 using Hamilton's equation and Lagrange's planetary equations}, one can use the following equation to determine the inclination evolution of the planets in the system over a given period of time.
\begin{equation}
    I_j(t) = \sqrt{p_j(t)^2 + q_j(t)^2} 
    \label{eq:inc_evo}
\end{equation}
    
These equations present the evolution of orbital parameters for each planet, dependant upon the stellar and planetary parameters. An example of what this time series looks like is shown in Figure \ref{fig:1e-04SecInc} for a $J_2 = 3.3 \cdot 10^{-6}$ and the initial parameters set to be the current-day values (Table \ref{tab:ParamsTable}). 
Figure \ref{fig:1e-04SecInc} illustrates the expected evolution based on the secular equations of motion, where the exact amplitudes of the motion are set by the masses and semi-major axes of the planets. Since we use second-order secular theory, the initial eccentricities do not factor into the time evolution of the inclination angle.

\subsection{Specific Predictions for TOI-125}
The goal of our secular analysis is to determine how the dynamics of the planetary systems responds to a change in stellar $J_{2}$, as the stellar $J_{2}$ is expected to decrease as the planetary system ages. 
To do this, we can use the equations outlined above to map how the dynamics of the system will change for different values of $J_{2}$. 

These equations were solved in Python for the parameters in Table \ref{tab:ParamsTable}, over a $10^{5}$ year integration, which was long enough to include at least several secular cycles for all values of $J_{2}$.
We were particularly interested in understanding the planets' secular behavior as related to the $J_2$ of the star, and so to test this, we computed the inclination evolution described by Equation \ref{eq:inc_evo} for a range of 500 discrete $J_2$ values equally spaced in log-space between between $10^{-2}$ and $10^{-9}$. These computations were identical except for the size of the $J_2$, as all planetary parameters were set to the be best-fit values from \citet{Quinn2019} and \citet{Nielsen2020}. The values used are reproduced in Table \ref{tab:ParamsTable}.  For each value of $J_2$, we compute the inclination evolution and generate a time series of inclination over time, each of which looks analogous to the time series plotted in Figure \ref{fig:1e-04SecInc} but with periods and amplitudes that depend on the exact value of $J_2$ used.

\begin{figure}
    \centering
    \includegraphics[width=3.4in]{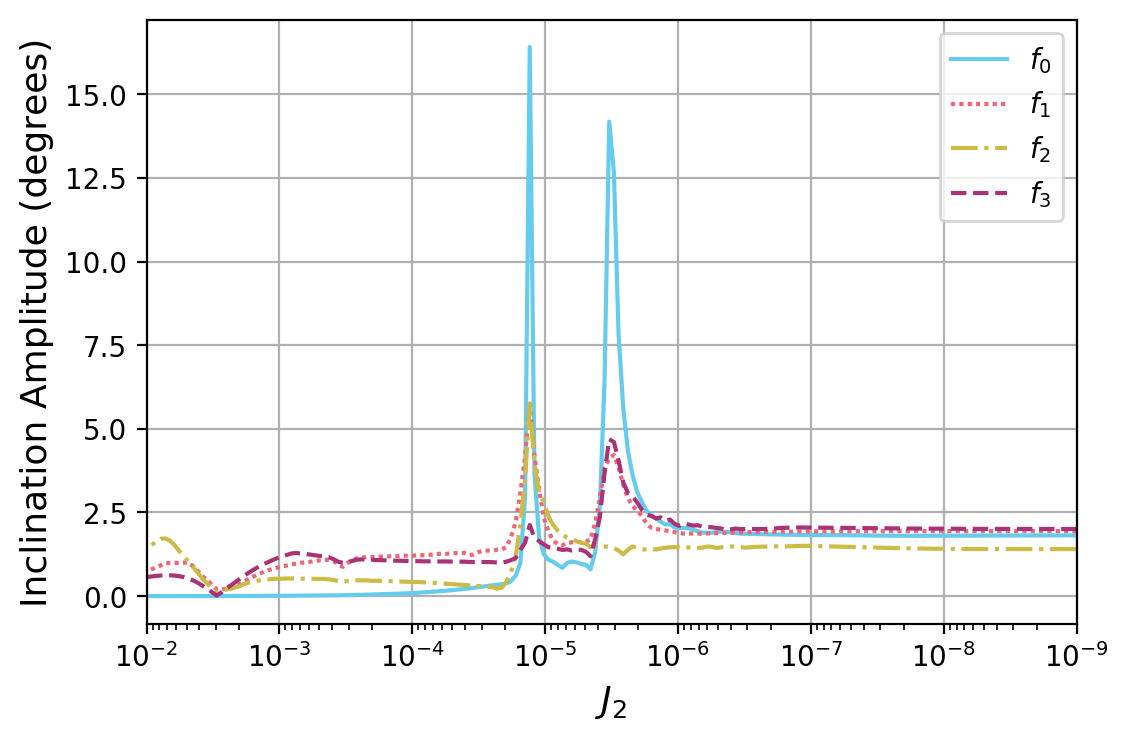}
    \caption{Plot of inclination amplitude as a function of $J_2$ for the TOI-125 planets using the secular theory. Amplitude in this case is the difference between maximum and minimum inclination in a $10^{4}$ year integration, and the USP planet is represented by the {blue line labeled $f_0$}. There are two clear spikes, occurring at $J_2$ = $1.3 \cdot 10^{-5}$ and $3.3 \cdot 10^{-6}$. These peaks allow us to determine that at these values of $J_2$ a system-wide instability occurs, which decouple the planets and cause them to oscillate widely in their inclination.}
    \label{fig:SecspikesvsJ2}
\end{figure}

For each discrete value of $J_2$, we found the respective {maxima} of the inclination oscillations. The value for each planet by $J_2$ is plotted in Figure \ref{fig:SecspikesvsJ2}. This figure shows two values of $J_2$ where the oscillation amplitudes reach values upwards of 5 degrees for the outer planets and more than 10 degrees for TOI-125.04. 
These spikes are atypical behavior, with oscillation amplitudes generally being much lower. This result demonstrates that for most values of $J_2$, the planetary dynamics will be very similar, but for a relatively small range of $J_2$ values, the dynamics will look very different. Figure \ref{fig:1e-04SecInc} demonstrates the behavior seen at one of these local maximums of dynamical activity. 

\begin{figure}
    \centering
    \includegraphics[width=3.4in]{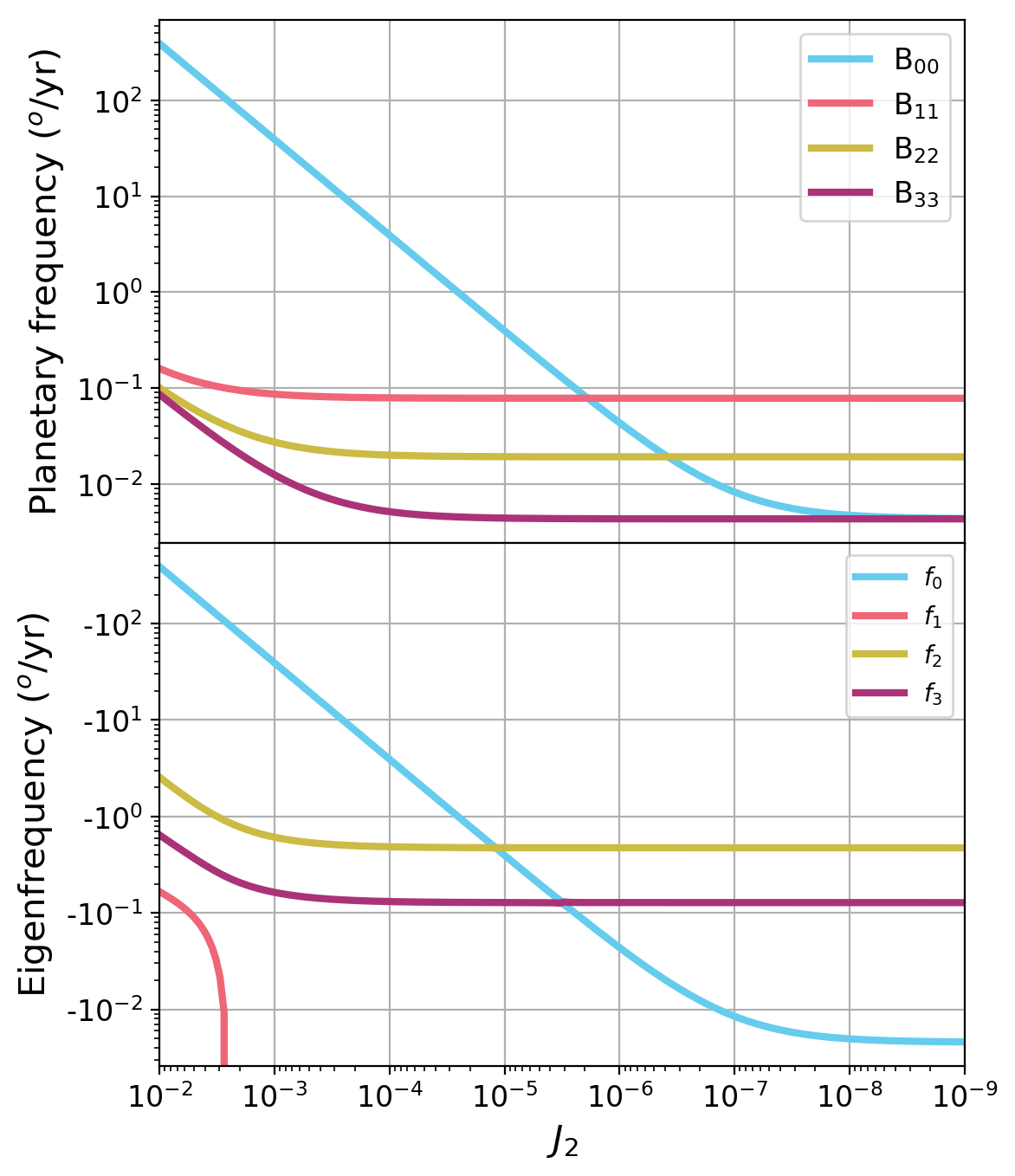}
    \caption{{Top panel: Plot of the intrinsic frequencies (Equation \ref{eq:intrinsicfreq}) for each planet. Bottom panel:} Plot of eigenfrequencies of {the system} as a function of $J_2$. A clear commensurability occurs between two of the four eigenfrequencies at two $J_2$ values: roughly $1.3 \cdot 10^{-5}$ and  $3.3 \cdot 10^{-6}$. Thus, an evolving $J_2$ (decreasing with time, as represented in the figure) can cause {dynamical action in the system as the system passes through these commensurabilities. Additionally, the intrinsic frequency for the inner planet, TOI-125.04 ($B_{00}$), is approximately equal to the eigenfrequency for that planet, meaning that those crossings correspond to linear secular resonances}.}
    \label{fig:EigfreqvsJ2}
\end{figure}

To understand the behavior shown in Figure \ref{fig:SecspikesvsJ2} more quantitatively, we can examine the modes of the dynamics that are described by the secular theory. In Figure \ref{fig:EigfreqvsJ2}, we plot the eigenfrequencies of the ${B}$ matrix as a function of $J_2$. Each frequency corresponds to a different {mode} driving the overall dynamics of the system. 

{The system eigenfrequencies and planetary intrinsic frequencies (given in Equation \ref{eq:intrinsicfreq}) together can be used to compute when resonances occur in the system. Secular resonances can lead to excitation of orbital elements, and occur when the eigenfrequencies of the system are commensurate with the planetary frequencies (or some combination of frequencies). When a intrinsic planetary frequency is commensurate with a system eigenfrequency, a linear secular resonance will occur \citep[ex:][]{Namouni1999,Batygin2008}, which can lead to large excitations in orbital elements.}

In a system where the host star's $J_2$ is decreasing with time, the system as a whole will pass through these points {of secular resonance, where a planetary intrinsic frequency becomes commensurate with a system eigenfrequency,} which will substantially alter the distribution of the angular momentum within the system.  The times (values of $J_2$) where this will happen are predicted by the secular theory, which can be used to compute at what values of $J_2$ the {relevant} frequencies will be close. For TOI-125, the specific values of $J_2$ at which the instabilities will occur is given by the intersections in Figure \ref{fig:EigfreqvsJ2}. {While the total angular momentum deficit (AMD) will generally be conserved \citep{Laskar2017}, its distribution among planets can be altered significantly at these points, {resulting in the relative values of eccentricity and inclination for each planet changing. This can be seen in the definition of AMD found in \citet{Laskar1997, Laskar2017}:
\begin{equation}
    C = \sum_{k=1}^n \Lambda_k (1-\sqrt{1-e^2_k}\cos{i_k})
\end{equation} 
Where $\Lambda_k = m_k\sqrt{\mu a_k}$ ($\mu$ reducing further into $GM$, where $G$ is the gravitational constant and $M$ is the mass of the star), $e_k$ is the eccentricity of the $k^{th}$ body, and $i_k$ is its accompanying inclination.} The points at which this will occur can approximately but not exactly be predicted by the secular theory. }

Since the results found in the theory described above can be perturbed by additional effects not present in the secular model \citep[ex:][]{Granados2018}, to further and more exactly examine how this evolution proceeds with time, we turn now to numerical simulations of the planetary system under the influence of the evolving star.

\begin{figure}[h!]
    \centering
    \includegraphics[width=3.4in]{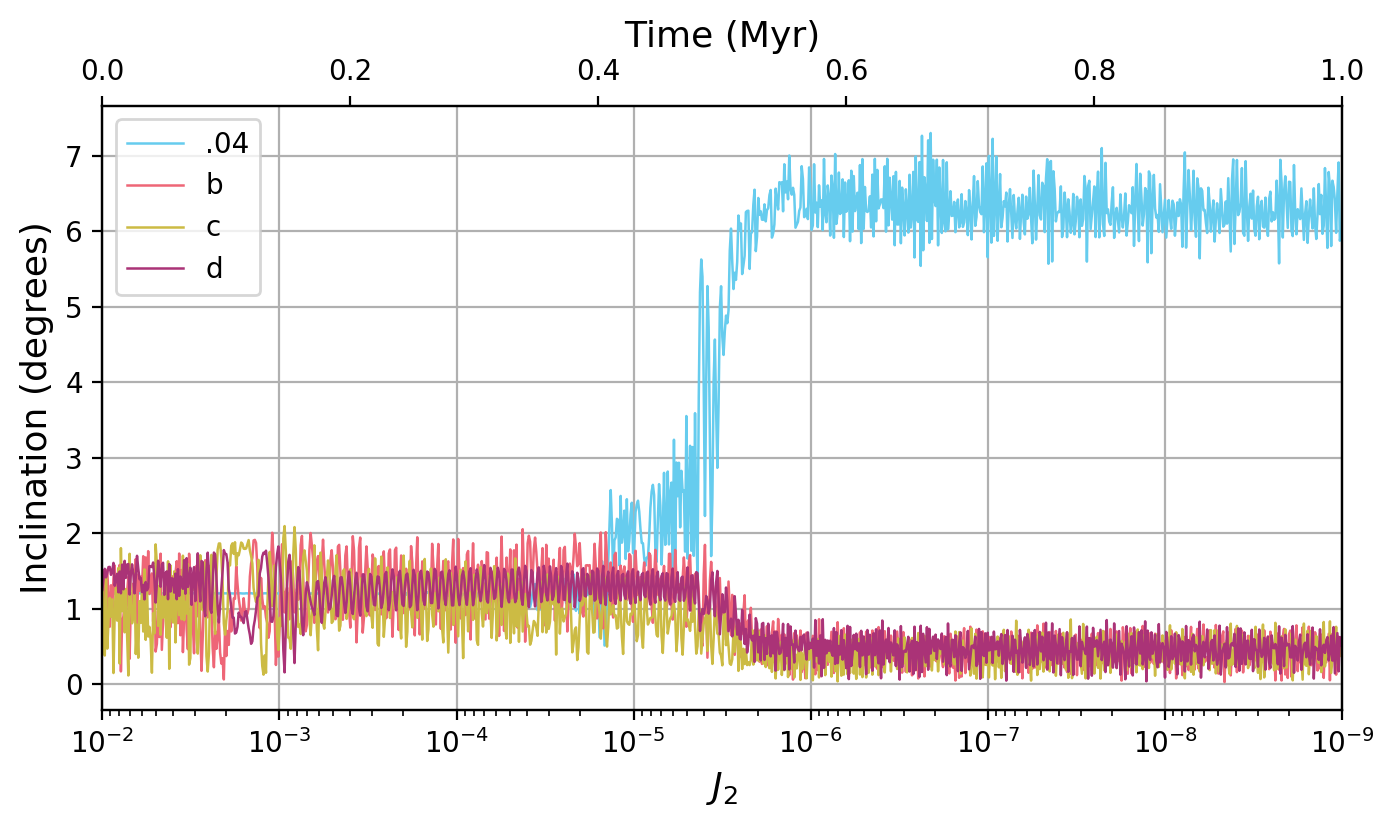}
    \caption{Plot of the inclination evolution of the four planets in TOI-125 as a function of $J_2$, for the 1 degree obliquity run. The USP planet is represented by the blue color, labeled .04, and shows a clear transition away from the original orbital plane as $J_2$ decreases.}
    \label{fig:1degIncvsJ2}
\end{figure}
\begin{figure}[h!]
    \centering
    \includegraphics[width=3.4in]{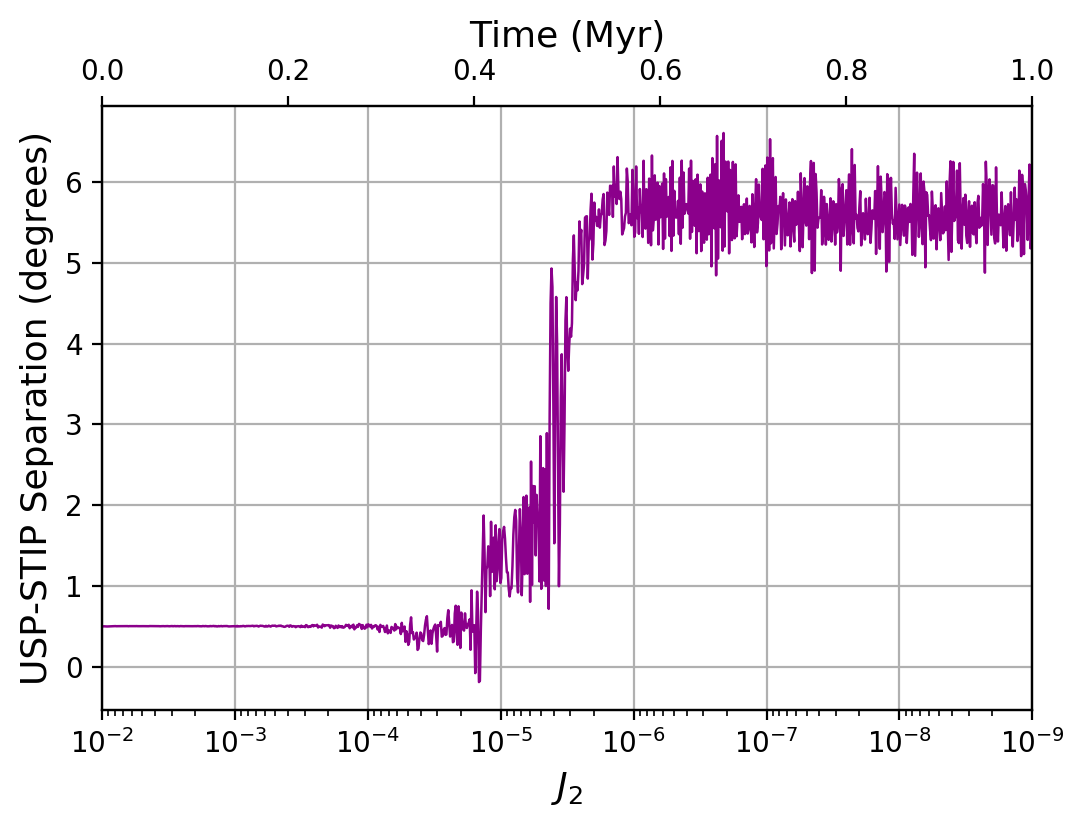}
    \caption{Plot of the USP-STIP misalignment in degrees as a function of $J_2$ for the 1 degree obliquity. This misalignment was calculated by taking the median of the STIP inclinations for each step of the loop and subtracting it from the inclination values of .04. It is thus clear to see the transition of the USP planet out of the orbital plane, due to the instabilities discussed in Figure \ref{fig:EigfreqvsJ2}.}
    \label{fig:1degUSPSTIPsep}
\end{figure}
\section{Numerical Simulations} 
\label{sec:nbody}
To study the long-term dynamical evolution of the TOI-125 system, we use N-body simulations, which allow for a full characterization of how the planetary orbits will change as $J_2$ evolves. 
The second order secular theory described above will solve for the inclination evolution decoupled from the eccentricity evolution, but the numerical integrations will compute the full solution and remove other approximations imposed by the secular theory, including the fixed semi-major axis ratios. 
To compute the N-body solutions, we use the Rebound \citep{Rein2012} and Reboundx \citep{Tamayo2020} packages.

\subsection{A Numerical Exploration of Long-Term Dynamics}
The purpose of the N-body integrations is to study the evolution of the known planets in the TOI-125 system as the star ages (and its $J_2$ moment decreases). The initial conditions of the planets are set to be a combination of the current-day values and some assumptions about what the system looked like at formation. 
At the time of disk dispersal, it is expected that the planets would have resided in a roughly coplanar configuration, so the initial planetary inclinations are confined to a narrow disk, being drawn from a Rayleigh distribution with a width of 1.4 degrees \citep{Fabrycky2014}. 
The star was also assigned an initial obliquity with respect to the mean plane containing the planets. The initial value tested for the stellar obliquity was 1 degree.  
Otherwise, the simulation was initialized using the planetary and stellar parameters reported in Table \ref{tab:ParamsTable}. 
This initial set-up corresponds to the following assumptions about the system's formation: we assume that the planets migrated into their final orbital radii in a coplanar manner, resulting in a initial condition that matches the orbital parameters summarized in \ref{tab:ParamsTable}. Some slight dispersion in initial inclination is allowed, but the mutual inclinations are low to start and allowed to change over the course of the integration. 

\begin{figure*}
    \centering
    \includegraphics[width=6in]{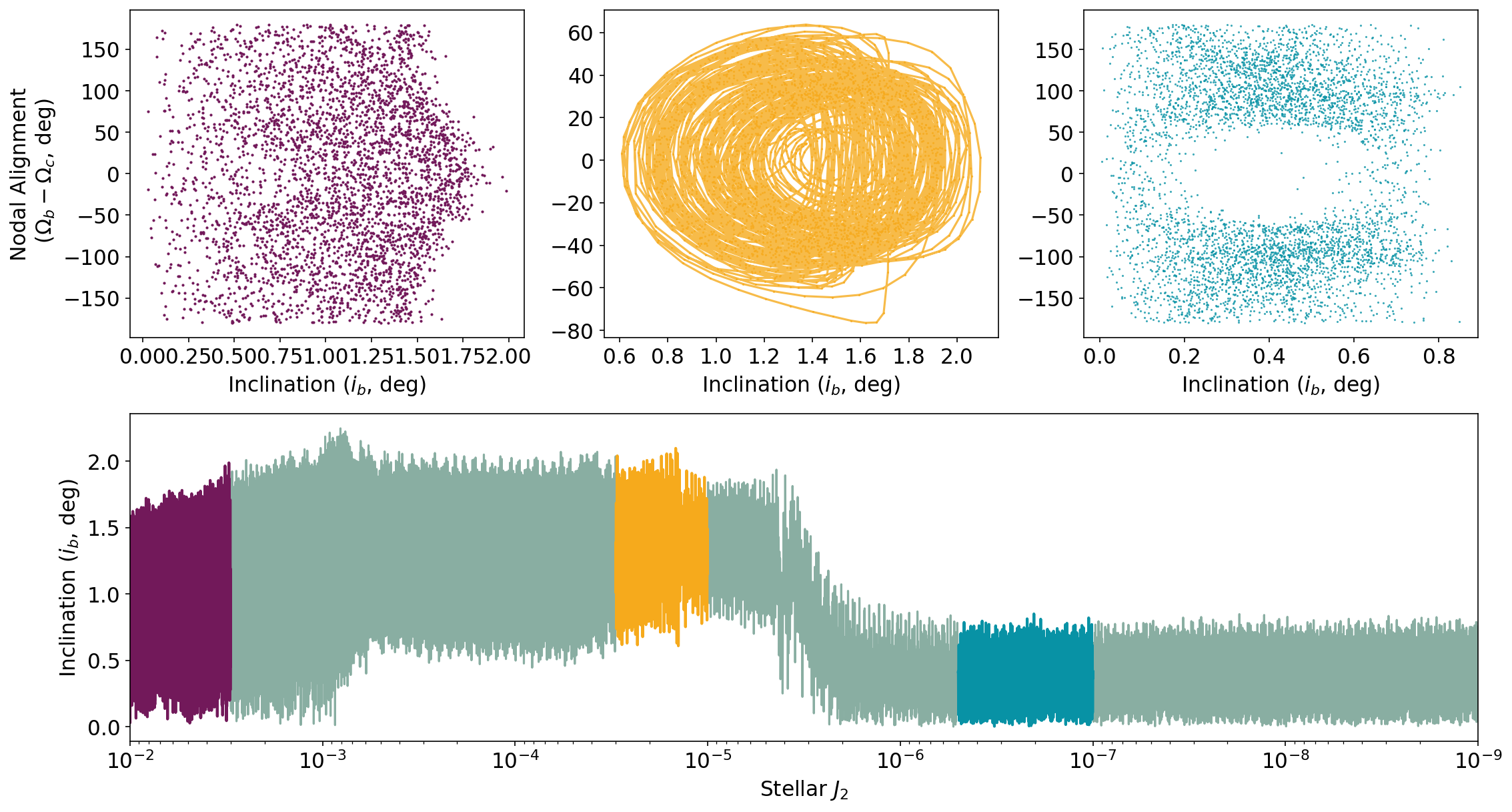}
    \caption{Figure representing the evolution of the nodal alignment of planet TOI-125b with TOI-125c as a function of inclination over the course of stellar spin down for a 1 Myr integration, in the 1 degree obliquity run. The three top phase portraits correspond with the $J_2$ ranges of the same color in the bottom plot, showing that as the system evolves, the phase spaces transition from circulation to libration. This implies that as the star spins down, the nodal alignment-inclination relationship becomes less oscillatory and more chaotic, a situation which could warrant further study.}
    \label{fig:my_label}
\end{figure*}
The simulated system was integrated using the \texttt{ias15} integrator. Each simulation was run for 1 Myr (roughly $7 \times 10^{9}$ orbits of the inner planet). {The effects of {general relativity} (GR) and the $J_2$ potential were included using the Reboundx \citep{Tamayo2020} package. To model $J_2$, we used the gravitational harmonics package\footnote{See https://reboundx.readthedocs.io/en/latest/effects.html and the gravitational harmonics example}, with the stellar axis fixed to the z-axis of the simulation. The $J_2$ of the star was initialized with a value of $10^{-2}$ at a time of 0, and decreased linearly in log-space over time, ending at $10^{-9}$ at 1 Myr of simulation time \citep{Batygin2013, Bouvier2013}. A $J_2 =  10^{-2}$ corresponds to a ratio of the stellar rotational frequency to breakup of around $\Omega_*/\Omega_{*,b} = 0.3$, which resides at roughly the 50th percentile for young stars according to \citet{Gallet2013}. }
Although the stellar quadrupole moment will decay much more slowly than our simulated rate over a real star's life, the dynamical time of the system is extremely short compared to the timescale of the change in $J_2$.
The integration time (1 Myr) was chosen to both allow computation feasibility of the problem (each integration as described above took one week on a single CPU core running at roughly 2.8 GHz), as well as ensure that the character of the dynamics is preserved as compared were the evolution to proceed for the full multi-Gyr system lifetime. 

The inclination evolution of the planets in an integration with the properties described above and for an initial stellar obliquity of one degree is shown in Figure \ref{fig:1degIncvsJ2}. 
Early in the simulation, where the $J_2$ is larger, all four planets remain in the same orbital plane, with only slight oscillations about the median; however, once the $J_2$ {decreases} to a value of $1.3\cdot 10^{-5}$, the planets reach the first predicted secular instability, at which time TOI-125.04 begins to become misaligned with the original orbital plane. At the same time, the three planets remaining within the plane also transition to a new mean plane location, except instead of their inclinations increasing with respect to the initial planar value, they decrease in order to conserve the AMD. This system-wide inclination evolution continues until the star spins down to a $J_2$ of $3.3\cdot10^{-6}$, where the final secular instability is encountered. The transition of TOI-125.04 out of the plane levels off at this $J_2$ value, and the STIP plane stabilizes in its inclination as well. After this point, the orbits of all four planets are relatively consistent as the star continues to spin down. 

Figure \ref{fig:1degUSPSTIPsep} plots the USP misalignment from the STIP plane for this particular initial stellar obliquity. The location of the STIP plane was defined as the median inclination of the outer three planets. 
Compared to the secular prediction in Figure \ref{fig:EigfreqvsJ2}, the large transition in the USP-STIP alignment occurs right where the secular theory predicts an instability. 
This integration suggests that for a very small initial stellar obliquity (1 degree), the USP-STIP misalignment will be measurable ($\sim7$ degrees) when the system is mature. {Note that for large differences in planetary inclinations, variations in $\Omega$ may also affect the observed alignment of the system, but for transiting systems as considered here the effect is very small.}

Simultaneously, as the systems evolves, the phase-space behavior of the planets will similarly change. In Figure  \ref{fig:my_label}, we show the both the inclination evolution for the second planet TOI-125 b (bottom panel) and the $i-\Omega$ phase space evolution for three representative times points (top panels). The angle $\Omega_{b} - \Omega_{c}$ is computed using the differential nodal alignment between adjacent planets, and this angle changes between libration and circulation as the $J_2$ decays and the planet transitions from an early nodal circulation (because the nodal precession caused by $J_2$ is large compared to the libration cycles in $\Omega$ induced by the planet-planet interactions) to later nodal libration (as the planet-planet effects dominate the dynamics). The point at which this transition between circulating and librating occurs differs for each planet, as it depends strongly on the planet's orbital period. The transition in the range $J_2 \sim 10^{-5} - 10^{-6}$ also denotes a transition between clear libration cycles between planets b and c and a more complex phase space where TOI-125.04 plays a more important role in the dynamics.

\subsection{The Effect of Varying the Initial Stellar Obliquity }
The integration presented in Figure \ref{fig:1degIncvsJ2} and Figure \ref{fig:1degUSPSTIPsep} demonstrates the behavior of the TOI-125 system with a star that has an initial obliquity of 1 degree relative to the plane of the planets.  
Such low obliquities could be generated easily if the precession rates of the planet-forming disk and the stellar axis have some very slight mismatch, and the approximate resultant mismatch between the inner and outer planet inclinations will be roughly 7 degrees, which is smaller than the current day value observed in the TOI-125 system.
\begin{figure}
    \centering
    \includegraphics[width=3.3in]{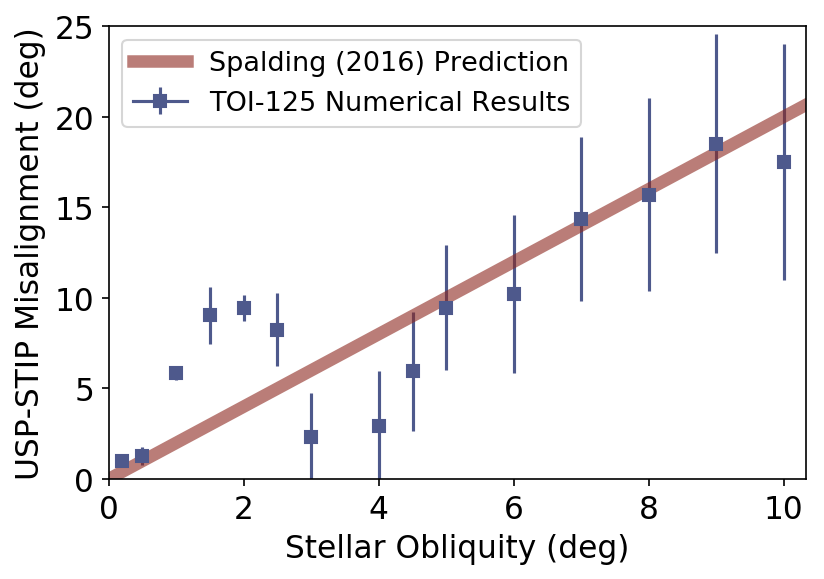}
    \caption{For a variety of initial stellar obliquities with respect to the plane containing the four TOI-125 planets, the eventual USP-STIP misalignment computed numerically after the star is allowed to evolve to $J_2 \sim 10^{-8}$. The prediction of \citet{Spalding2016} for a two-planet system is overlaid. The multi-planet dynamics in this case is more complex than the prediction, but the prediction does a good job of predicting the general trend of USP-STIP misalignment. }
    \label{fig:obliquity}
\end{figure}
To check how the misalignment at late times will be affected by the initial value of the stellar obliquity, we performed a total of 15 integrations with identical parameters as described in the previous section, except for different initial values of the stellar obliquity between 0.2 and 10 degrees. 
For each integration, we compute the median and spread in the USP-STIP misalignment at late times (when $J_2 > 10^{-7}$). The results are plotted in Figure \ref{fig:obliquity}. The errorbars indicate the distribution of misalignment values present within one integration, as the planetary inclinations will oscillate with time. 

As predicted in \citet{Spalding2016}, the misalignment between adjacent planets increases for an increased initial stellar obliquity. In Figure \ref{fig:obliquity}, we also plot the prediction for a simple two-planet system from \citet{Spalding2016} for small value of $J_2$ < $10^{-6}$. The prediction for such a simplified system is that the misalignment between adjacent planets in a two-planet system (measured in degrees of inclination) will be twice the stellar obliquity. The numerically computed values of the mutual planet inclinations (between the USP planet and the STIP) generally follow this trend, with deviations at low values of stellar obliquity. The deviation between the numerical results for TOI-125 and the analytic prediction of \citet{Spalding2016} is explained by the complexity of multi-planet dynamics: rather then a two-planet pair, the TOI-125 system has a set of three planets exterior to the USP planet. As shown in Figure \ref{fig:my_label}, the phase space behavior of two planets in the STIP changes over time. At later times where STIP components no longer circulate in $\Omega_{b} - \Omega_{c}$, this planet-planet coupling can significantly alter the planets' precession frequencies.
The presence of additional planets beyond the two-planet system considered in \citet{Spalding2016} serves to perturb the precession frequency of TOI-125 b, resulting in the presence of secular resonances at particular values of $J_2$ and deviation between the model and simulation results presented in Figure \ref{fig:obliquity}. We also note that for much larger stellar obliquities than are considered here, the multi-planet dynamics may cause further deviation from the theory.

\section{Discussion}
\label{sec:discuss}
The Kepler data set yielded constraints on exoplanet demographics which suggested that tightly packed, multi-planet systems are relatively common \citep{Muirhead2015}, despite their dissimilarity to our own solar system. 
Although exoplanet system geometries with misalignment in the inclination angle between components of a STIP apparently subvert this archetype, simultaneously, in the exoplanet sample as a whole, planets with particularly short orbital periods in multi-planet systems have a larger range of inclinations compared to other planets in their systems \citep{Dai2018}. 
Previous work has discussed how this orbital-radius-dependant inclination dispersion can be related to stellar parameters, including the stellar quadrupole moment and stellar obliquity \citep{Li2020, Becker2020}.
In this work, we have studied a system, TOI-125, whose tightly-packed geometry fits into this picture, with a high mutual inclination for the ultra-short-period planet, and a coplanar geometry for the outer planets. We use this system as a test case, and expand on previous work to demonstrate how an evolving star can naturally generate the system geometry described above from a typical, multi-planet system that formed in the protoplanetary disk. {In our analysis, we assume that the disk dissipates sufficiently rapidly that the planet-planet dynamics is allowed to emerge by the time the secular resonance crossings occur, and we also assume that the disk dissipation is fast enough that the natal misalignment between the star and the disk is retained as star-planet misalignment \citep{Spalding2020}.} We also show that the specific initial stellar parameters (particularly the stellar obliquity) set the exact geometry (the USP-STIP angle) which will be observed after the system has evolved to a relatively old age.

This evolution in planetary inclination is born from asymmetry in the stellar potential, parameterized both as a stellar quadrupole moment ($J_2$) and a stellar obliquity. 
If we assume that a system formed in a protoplanetary disk, some misalignment between the host star and the planet-forming disk must exist in order for the dynamics described above to become relevant. 
Otherwise, as shown for the low values of stellar obliquity in Figure \ref{fig:obliquity}, the USP-STIP misalignment will remain low. 
A stellar obliquity can originate in, for example, the presence of some misaligned external potential, either stellar \citep{Spalding2014} or due to a misaligned outer component of the protoplanetary disk \citep{Loomis2017, Ansdell2020, Francis2020}. 

Considering the evolving system from a secular perspective, the relative importance of the planetary eigenfrequencies in the system will change as the stellar quadrupole moment decays. At times where fundamental system frequencies are very close to each other and planetary frequencies, this can cause {the emergence of secular resonances that were not present at other $J_2$ values. This effect can cause a redistribution of the phase space area of the system while conserving the total AMD \citep[see also][]{Faridani2021}.}
This is seen clearly in Figure \ref{fig:1degIncvsJ2}, where the increase in the inner planet's orbital inclination occurs simultaneously with the alignment of the outer planets with the stellar equator. The AMD of the system as a whole is conserved, and for the dynamics in this paper, the evolution takes place at constant semi-major axis. 
\citet{Petrovich2018} describes a similar chaotic instability due to secular resonances {and tidal effects} which can result in significant migration of planets, but the dynamics described in this paper is comparatively non-violent and will lead only to a redistribution of AMD rather than planetary migration in semi-major axis.

It is important to note that the locations of these instabilities will change for different geometries of systems. In this paper, we have considered the TOI-125 system, and during its evolution two instabilities occurred (as shown in Figure \ref{fig:EigfreqvsJ2}). 
In the TOI-125 system, the current day misalignment between the USP planet and the outer three planets is 14 - 18 degrees. Based on the results of Figure \ref{fig:obliquity}, this corresponds to an initial $\sim 5-10 $ degree stellar obliquity with respect to the forming plane of planets. {This value is similar to that of the solar obliquity \citep{Beck2005}.}

In other systems with differing planet multiplicities or planet parameters, these instabilities may be more or less common. 
In many cases, they may not occur at all, both due to the fundamental system parameters as well as if the planets arrive in their final orbital positions after the star has spun down sufficiently. 
Systems will attain the geometry described in this paper when the planets arrive in their final orbital locations at a time before that when the planetary fundamental frequencies will become commensurate. This occurs in the case where there exists some non-zero stellar obliquity with respect to the inner planet forming disk.
Even for a relatively small initial stellar obliquity, the USP-STIP misalignment will be measurable when the system is a few Gyr old. 
However, in systems where the stellar obliquity is sufficiently low or in which planets arrive close to the star after the quadrupole moment has decayed significantly for the planets' particular orbital radii, then the inclination dynamics described in this paper will not arise.

For this latter set of systems, limits on dynamical stellar obliquity or the arrival time of planets (if the true obliquity history is known) could be possible. In the TOI-125 system, the planets likely arrived into their observed locations early enough that the dynamics in this paper was allowed to operate. {If the planets migrated inwards too late, then $J_2$ would have decayed sufficiently such that the inclination-exciting dynamics would no longer operate; see Figure \ref{fig:my_label}, which shows that for this particular geometry, the dynamics no longer actively alters orbits after a $J_2$ of $10^{-6}$ or so \citep[see also discussion in ][]{Schultz2021}.}
The mechanism described in this paper will not occur for planets will sufficiently large orbital radii. 

\section{Conclusion} 
\label{sec:conclude}
 
We present the relationship between the observed misalignment between TOI-125.04 and the outer three planets in the TOI-125 system as it depends on the stellar quadrupole moment and the stellar obliquity. 
Using the secular perturbation theory, we were able to predict values of $J_2$ at which the TOI-125 system would be expected to have instabilities in inclination angle (Figure \ref{fig:SecspikesvsJ2}). We then evaluated the long-term evolution of the system with changing $J_2$ using N-body integrations, in which we showed that the conservation of the phase space area results in a redistribution of the AMD in the system, creating a misalignment between TOI-125.04 and the other planets which emerges at roughly the time predicted by the secular theory. 
We conclude that if TOI-125.04 had started in the same orbital plane as the other planets in the TOI-125 system and if the system had some small ($\sim 5-10 $ degree) stellar obliquity, TOI-125.04 could naturally have migrated to its current 14 degree separation as a product of the stellar spin down.
{The mechanism described in this paper can be applied more generally to systems with STIPs and inner misaligned USP planets, and naturally explains their geometries as observed at present day.}

\medskip
\textbf{Acknowledgements.}
We would like to thank Thea Faridani for useful conversations, and Konstantin Batygin for guidance and advice.  J.C.B.~has been supported by the Heising-Simons \textit{51 Pegasi b} postdoctoral fellowship. L.B. also thanks the Heising-Simons foundation for financial support. We would also like to thank the UROP / Research Scholars program at the University of Michigan for providing this research opportunity and for the continued support throughout these difficult times. We would like to thank the anonymous referee for useful comments that improved the manuscript.

Software: pandas \citep{mckinney-proc-scipy-2010}, IPython \citep{PER-GRA:2007}, matplotlib \citep{Hunter:2007}, scipy \citep{scipy}, numpy \citep{oliphant-2006-guide}, Jupyter \citep{Kluyver:2016aa}, Rebound \citep{Rein2012}, Reboundx \citep{Tamayo2020}

\bibliography{refs}

\begin{thebibliography}{}
\expandafter\ifx\csname natexlab\endcsname\relax\def\natexlab#1{#1}\fi
\providecommand{\url}[1]{\href{#1}{#1}}
\providecommand{\dodoi}[1]{doi:~\href{http://doi.org/#1}{\nolinkurl{#1}}}
\providecommand{\doeprint}[1]{\href{http://ascl.net/#1}{\nolinkurl{http://ascl.net/#1}}}
\providecommand{\doarXiv}[1]{\href{https://arxiv.org/abs/#1}{\nolinkurl{https://arxiv.org/abs/#1}}}

\bibitem[{{Adams} {et~al.}(2021){Adams}, {Jackson}, {Johnson}, {Ciardi},
  {Cochran}, {Endl}, {Everett}, {Furlan}, {Howell}, {Jayanthi}, {MacQueen},
  {Matson}, {Partyka-Worley}, {Schlieder}, {Scott}, {Stanton}, \&
  {Ziegler}}]{Adams2020}
{Adams}, E.~R., {Jackson}, B., {Johnson}, S., {et~al.} 2021, Planetary Science
  Journal, 2, 152, \dodoi{10.3847/PSJ/ac0ea0}

\bibitem[{{Ansdell} {et~al.}(2020){Ansdell}, {Gaidos}, {Hedges}, {Tazzari},
  {Kraus}, {Wyatt}, {Kennedy}, {Williams}, {Mann}, {Angelo}, {D{\^u}chene},
  {Mamajek}, {Carpenter}, {Esplin}, \& {Rizzuto}}]{Ansdell2020}
{Ansdell}, M., {Gaidos}, E., {Hedges}, C., {et~al.} 2020, \mnras, 492, 572,
  \dodoi{10.1093/mnras/stz3361}

\bibitem[{{Batygin}(2015)}]{Batygin2015}
{Batygin}, K. 2015, \mnras, 451, 2589, \dodoi{10.1093/mnras/stv1063}

\bibitem[{{Batygin} \& {Adams}(2013)}]{Batygin2013}
{Batygin}, K., \& {Adams}, F.~C. 2013, \apj, 778, 169,
  \dodoi{10.1088/0004-637X/778/2/169}

\bibitem[{{Batygin} \& {Laughlin}(2008)}]{Batygin2008}
{Batygin}, K., \& {Laughlin}, G. 2008, \apj, 683, 1207, \dodoi{10.1086/589232}

\bibitem[{{Beck} \& {Giles}(2005)}]{Beck2005}
{Beck}, J.~G., \& {Giles}, P. 2005, \apjl, 621, L153, \dodoi{10.1086/429224}

\bibitem[{{Becker} {et~al.}(2021){Becker}, {Batygin}, \& {Adams}}]{Becker2021}
{Becker}, J., {Batygin}, K., \& {Adams}, F. 2021, arXiv e-prints,
  arXiv:2107.03413.
\newblock \doarXiv{2107.03413}

\bibitem[{{Becker} {et~al.}(2020){Becker}, {Batygin}, {Fabrycky}, {Adams},
  {Li}, {Vanderburg}, \& {Rodriguez}}]{Becker2020}
{Becker}, J., {Batygin}, K., {Fabrycky}, D., {et~al.} 2020, \aj, 160, 254,
  \dodoi{10.3847/1538-3881/abbad3}

\bibitem[{{Becker} {et~al.}(2015){Becker}, {Vanderburg}, {Adams}, {Rappaport},
  \& {Schwengeler}}]{Becker2015}
{Becker}, J.~C., {Vanderburg}, A., {Adams}, F.~C., {Rappaport}, S.~A., \&
  {Schwengeler}, H.~M. 2015, \apjl, 812, L18,
  \dodoi{10.1088/2041-8205/812/2/L18}

\bibitem[{{Bouvier}(2013)}]{Bouvier2013}
{Bouvier}, J. 2013, in EAS Publications Series, Vol.~62, EAS Publications
  Series, ed. P.~{Hennebelle} \& C.~{Charbonnel}, 143--168,
  \dodoi{10.1051/eas/1362005}

\bibitem[{{Dai} {et~al.}(2018){Dai}, {Masuda}, \& {Winn}}]{Dai2018}
{Dai}, F., {Masuda}, K., \& {Winn}, J.~N. 2018, \apjl, 864, L38,
  \dodoi{10.3847/2041-8213/aadd4f}

\bibitem[{{Deck} \& {Batygin}(2015)}]{Deck2015}
{Deck}, K.~M., \& {Batygin}, K. 2015, \apj, 810, 119,
  \dodoi{10.1088/0004-637X/810/2/119}

\bibitem[{{Fabrycky} {et~al.}(2014){Fabrycky}, {Lissauer}, {Ragozzine}, {Rowe},
  {Steffen}, {Agol}, {Barclay}, {Batalha}, {Borucki}, {Ciardi}, {Ford},
  {Gautier}, {Geary}, {Holman}, {Jenkins}, {Li}, {Morehead}, {Morris},
  {Shporer}, {Smith}, {Still}, \& {Van Cleve}}]{Fabrycky2014}
{Fabrycky}, D.~C., {Lissauer}, J.~J., {Ragozzine}, D., {et~al.} 2014, \apj,
  790, 146, \dodoi{10.1088/0004-637X/790/2/146}

\bibitem[{{Faridani} {et~al.}(2021){Faridani}, {Naoz}, {Wei}, \&
  {Farr}}]{Faridani2021}
{Faridani}, T., {Naoz}, S., {Wei}, L., \& {Farr}, W.~M. 2021, arXiv e-prints,
  arXiv:2107.07529.
\newblock \doarXiv{2107.07529}

\bibitem[{{Francis} \& {van der Marel}(2020)}]{Francis2020}
{Francis}, L., \& {van der Marel}, N. 2020, \apj, 892, 111,
  \dodoi{10.3847/1538-4357/ab7b63}

\bibitem[{{Gallet} \& {Bouvier}(2013)}]{Gallet2013}
{Gallet}, F., \& {Bouvier}, J. 2013, \aap, 556, A36,
  \dodoi{10.1051/0004-6361/201321302}

\bibitem[{{Granados Contreras} \& {Boley}(2018)}]{Granados2018}
{Granados Contreras}, A.~P., \& {Boley}, A.~C. 2018, \aj, 155, 139,
  \dodoi{10.3847/1538-3881/aaac82}

\bibitem[{Hunter(2007)}]{Hunter:2007}
Hunter, J.~D. 2007, Computing In Science \& Engineering, 9, 90,
  \dodoi{10.1109/MCSE.2007.55}

\bibitem[{Jones {et~al.}(2001)Jones, Oliphant, Peterson, {et~al.}}]{scipy}
Jones, E., Oliphant, T., Peterson, P., {et~al.} 2001, {SciPy}: Open source
  scientific tools for {Python}.
\newblock \url{http://www.scipy.org/}

\bibitem[{Kluyver {et~al.}(2016)Kluyver, Ragan-Kelley, P{\'e}rez, Granger,
  Bussonnier, Frederic, Kelley, Hamrick, Grout, Corlay, Ivanov, Avila, Abdalla,
  \& Willing}]{Kluyver:2016aa}
Kluyver, T., Ragan-Kelley, B., P{\'e}rez, F., {et~al.} 2016, in Positioning and
  Power in Academic Publishing: Players, Agents and Agendas, ed. F.~Loizides \&
  B.~Schmidt, IOS Press, 87 -- 90

\bibitem[{{Laskar}(1997)}]{Laskar1997}
{Laskar}, J. 1997, \aap, 317, L75

\bibitem[{{Laskar} \& {Petit}(2017)}]{Laskar2017}
{Laskar}, J., \& {Petit}, A.~C. 2017, \aap, 605, A72,
  \dodoi{10.1051/0004-6361/201630022}

\bibitem[{{Li} {et~al.}(2020){Li}, {Dai}, \& {Becker}}]{Li2020}
{Li}, G., {Dai}, F., \& {Becker}, J. 2020, \apjl, 890, L31,
  \dodoi{10.3847/2041-8213/ab72f4}

\bibitem[{{Loomis} {et~al.}(2017){Loomis}, {{\"O}berg}, {Andrews}, \&
  {MacGregor}}]{Loomis2017}
{Loomis}, R.~A., {{\"O}berg}, K.~I., {Andrews}, S.~M., \& {MacGregor}, M.~A.
  2017, \apj, 840, 23, \dodoi{10.3847/1538-4357/aa6c63}

\bibitem[{{MacDonald} {et~al.}(2016){MacDonald}, {Ragozzine}, {Fabrycky},
  {Ford}, {Holman}, {Isaacson}, {Lissauer}, {Lopez}, {Mazeh}, {Rogers}, {Rowe},
  {Steffen}, \& {Torres}}]{MacDonald2016}
{MacDonald}, M.~G., {Ragozzine}, D., {Fabrycky}, D.~C., {et~al.} 2016, \aj,
  152, 105, \dodoi{10.3847/0004-6256/152/4/105}

\bibitem[{McKinney(2010)}]{mckinney-proc-scipy-2010}
McKinney, W. 2010, in Proceedings of the 9th Python in Science Conference, ed.
  S.~van~der Walt \& J.~Millman, 51 -- 56

\bibitem[{{Millholland} {et~al.}(2017){Millholland}, {Wang}, \&
  {Laughlin}}]{Millholland2017}
{Millholland}, S., {Wang}, S., \& {Laughlin}, G. 2017, \apjl, 849, L33,
  \dodoi{10.3847/2041-8213/aa9714}

\bibitem[{{Millholland} \& {Spalding}(2020)}]{Millholland2020}
{Millholland}, S.~C., \& {Spalding}, C. 2020, \apj, 905, 71,
  \dodoi{10.3847/1538-4357/abc4e5}

\bibitem[{{Muirhead} {et~al.}(2015){Muirhead}, {Mann}, {Vanderburg}, {Morton},
  {Kraus}, {Ireland}, {Swift}, {Feiden}, {Gaidos}, \& {Gazak}}]{Muirhead2015}
{Muirhead}, P.~S., {Mann}, A.~W., {Vanderburg}, A., {et~al.} 2015, \apj, 801,
  18, \dodoi{10.1088/0004-637X/801/1/18}

\bibitem[{{Murray} \& {Dermott}(1999)}]{MD99}
{Murray}, C.~D., \& {Dermott}, S.~F. 1999, {Solar system dynamics}

\bibitem[{{Namouni} \& {Murray}(1999)}]{Namouni1999}
{Namouni}, F., \& {Murray}, C.~D. 1999, \aj, 117, 2561, \dodoi{10.1086/300849}

\bibitem[{{Nielsen} {et~al.}(2020){Nielsen}, {Gandolfi}, {Armstrong},
  {Jenkins}, {Fridlund}, {Santos}, {Dai}, {Adibekyan}, {Luque}, {Steffen},
  {Esposito}, {Meru}, {Sabotta}, {Bolmont}, {Kossakowski}, {Otegi}, {Murgas},
  {Stalport}, {Rodler}, {D{\'\i}az}, {Kurtovic}, {Ricker}, {Vanderspek},
  {Latham}, {Seager}, {Winn}, {Jenkins}, {Allart}, {Almenara}, {Barrado},
  {Barros}, {Bayliss}, {Berdi{\~n}as}, {Boisse}, {Bouchy}, {Boyd}, {Brown},
  {Bryant}, {Burke}, {Cochran}, {Cooke}, {Demangeon}, {D{\'\i}az}, {Dittman},
  {Dorn}, {Dumusque}, {Garc{\'\i}a}, {Gonz{\'a}lez-Cuesta}, {Grziwa},
  {Georgieva}, {Guerrero}, {Hatzes}, {Helled}, {Henze}, {Hojjatpanah}, {Korth},
  {Lam}, {Lillo-Box}, {Lopez}, {Livingston}, {Mathur}, {Mousis}, {Narita},
  {Osborn}, {Palle}, {Rojas}, {Persson}, {Quinn}, {Rauer}, {Redfield},
  {Santerne}, {dos Santos}, {Seidel}, {Sousa}, {Ting}, {Turbet}, {Udry},
  {Vanderburg}, {Van Eylen}, {Vines}, {Wheatley}, \& {Wilson}}]{Nielsen2020}
{Nielsen}, L.~D., {Gandolfi}, D., {Armstrong}, D.~J., {et~al.} 2020, \mnras,
  492, 5399, \dodoi{10.1093/mnras/staa197}

\bibitem[{Oliphant(2006)}]{oliphant-2006-guide}
Oliphant, T.~E. 2006, Guide to NumPy, Provo, UT.
\newblock \url{http://www.tramy.us/}

\bibitem[{P\'erez \& Granger(2007)}]{PER-GRA:2007}
P\'erez, F., \& Granger, B.~E. 2007, Computing in Science and Engineering, 9,
  21, \dodoi{10.1109/MCSE.2007.53}

\bibitem[{{Petrovich} {et~al.}(2019){Petrovich}, {Deibert}, \&
  {Wu}}]{Petrovich2018}
{Petrovich}, C., {Deibert}, E., \& {Wu}, Y. 2019, \aj, 157, 180,
  \dodoi{10.3847/1538-3881/ab0e0a}

\bibitem[{{Pu} \& {Lai}(2019)}]{Pu2019}
{Pu}, B., \& {Lai}, D. 2019, \mnras, 488, 3568, \dodoi{10.1093/mnras/stz1817}

\bibitem[{{Quinn} {et~al.}(2019){Quinn}, {Becker}, {Rodriguez}, {Hadden},
  {Huang}, {Morton}, {Adams}, {Armstrong}, {Eastman}, {Horner}, {Kane},
  {Lissauer}, {Twicken}, {Vanderburg}, {Wittenmyer}, {Ricker}, {Vanderspek},
  {Latham}, {Seager}, {Winn}, {Jenkins}, {Agol}, {Barkaoui}, {Beichman},
  {Bouchy}, {Bouma}, {Burdanov}, {Campbell}, {Carlino}, {Cartwright},
  {Charbonneau}, {Christiansen}, {Ciardi}, {Collins}, {Collins}, {Conti},
  {Crossfield}, {Daylan}, {Dittmann}, {Doty}, {Dragomir}, {Ducrot}, {Gillon},
  {Glidden}, {Goeke}, {Gonzales}, {He{\l}miniak}, {Horch}, {Howell}, {Jehin},
  {Jensen}, {Kielkopf}, {Kristiansen}, {Law}, {Mann}, {Marmier}, {Matson},
  {Matthews}, {Mazeh}, {Mori}, {Murgas}, {Murray}, {Narita}, {Nielsen},
  {Ottoni}, {Palle}, {Paw{\l}aszek}, {Pepe}, {Pitogo de Leon}, {Pozuelos},
  {Relles}, {Schlieder}, {Sebastian}, {S{\'e}gransan}, {Shporer}, {Stassun},
  {Tamura}, {Udry}, {Waite}, {Winters}, \& {Ziegler}}]{Quinn2019}
{Quinn}, S.~N., {Becker}, J.~C., {Rodriguez}, J.~E., {et~al.} 2019, \aj, 158,
  177, \dodoi{10.3847/1538-3881/ab3f2b}

\bibitem[{{Rein}(2012)}]{Rein2012}
{Rein}, H. 2012, \mnras, 427, L21, \dodoi{10.1111/j.1745-3933.2012.01337.x}

\bibitem[{{Rodriguez} {et~al.}(2018){Rodriguez}, {Becker}, {Eastman}, {Hadden},
  {Vanderburg}, {Khain}, {Quinn}, {Mayo}, {Dressing}, {Schlieder}, {Ciardi},
  {Latham}, {Rappaport}, {Adams}, {Berlind}, {Bieryla}, {Calkins}, {Esquerdo},
  {Kristiansen}, {Omohundro}, {Schwengeler}, {Stassun}, \&
  {Terentev}}]{Rodriguez2018}
{Rodriguez}, J.~E., {Becker}, J.~C., {Eastman}, J.~D., {et~al.} 2018, \aj, 156,
  245, \dodoi{10.3847/1538-3881/aae530}

\bibitem[{{Sanchis-Ojeda} {et~al.}(2014){Sanchis-Ojeda}, {Rappaport}, {Winn},
  {Kotson}, {Levine}, \& {El Mellah}}]{SanchisOjeda2014}
{Sanchis-Ojeda}, R., {Rappaport}, S., {Winn}, J.~N., {et~al.} 2014, \apj, 787,
  47, \dodoi{10.1088/0004-637X/787/1/47}

\bibitem[{{Schultz} {et~al.}(2021){Schultz}, {Spalding}, \&
  {Batygin}}]{Schultz2021}
{Schultz}, K., {Spalding}, C., \& {Batygin}, K. 2021, \mnras, 506, 2999,
  \dodoi{10.1093/mnras/stab1899}

\bibitem[{{Spalding} \& {Batygin}(2014)}]{Spalding2014}
{Spalding}, C., \& {Batygin}, K. 2014, \apj, 790, 42,
  \dodoi{10.1088/0004-637X/790/1/42}

\bibitem[{{Spalding} \& {Batygin}(2016)}]{Spalding2016}
---. 2016, \apj, 830, 5, \dodoi{10.3847/0004-637X/830/1/5}

\bibitem[{{Spalding} \& {Millholland}(2020)}]{Spalding2020}
{Spalding}, C., \& {Millholland}, S.~C. 2020, \aj, 160, 105,
  \dodoi{10.3847/1538-3881/aba629}

\bibitem[{{Sterne}(1939)}]{Sterne1939}
{Sterne}, T.~E. 1939, \mnras, 99, 451, \dodoi{10.1093/mnras/99.5.451}

\bibitem[{{Swift} {et~al.}(2013){Swift}, {Johnson}, {Morton}, {Crepp},
  {Montet}, {Fabrycky}, \& {Muirhead}}]{Swift2013}
{Swift}, J.~J., {Johnson}, J.~A., {Morton}, T.~D., {et~al.} 2013, \apj, 764,
  105, \dodoi{10.1088/0004-637X/764/1/105}

\bibitem[{{Tamayo} {et~al.}(2020){Tamayo}, {Rein}, {Shi}, \&
  {Hernandez}}]{Tamayo2020}
{Tamayo}, D., {Rein}, H., {Shi}, P., \& {Hernandez}, D.~M. 2020, \mnras, 491,
  2885, \dodoi{10.1093/mnras/stz2870}

\bibitem[{{Ward} {et~al.}(1976){Ward}, {Colombo}, \& {Franklin}}]{Ward1976}
{Ward}, W.~R., {Colombo}, G., \& {Franklin}, F.~A. 1976, \icarus, 28, 441,
  \dodoi{10.1016/0019-1035(76)90117-2}

\bibitem[{{Weiss} {et~al.}(2018){Weiss}, {Marcy}, {Petigura}, {Fulton},
  {Howard}, {Winn}, {Isaacson}, {Morton}, {Hirsch}, {Sinukoff}, {Cumming},
  {Hebb}, \& {Cargile}}]{Weiss2018}
{Weiss}, L.~M., {Marcy}, G.~W., {Petigura}, E.~A., {et~al.} 2018, \aj, 155, 48,
  \dodoi{10.3847/1538-3881/aa9ff6}

\bibitem[{{Winn} {et~al.}(2018){Winn}, {Sanchis-Ojeda}, \&
  {Rappaport}}]{Winn2018}
{Winn}, J.~N., {Sanchis-Ojeda}, R., \& {Rappaport}, S. 2018, \nar, 83, 37,
  \dodoi{10.1016/j.newar.2019.03.006}

\bibitem[{{Zhu} {et~al.}(2018){Zhu}, {Petrovich}, {Wu}, {Dong}, \&
  {Xie}}]{Zhu2018}
{Zhu}, W., {Petrovich}, C., {Wu}, Y., {Dong}, S., \& {Xie}, J. 2018, \apj, 860,
  101, \dodoi{10.3847/1538-4357/aac6d5}

\end{thebibliography}

\begin{table*}[h!]
    \centering
    \begin{tabular}{ccccc}
         \hline
       {Stellar Parameters}  &  &  &  &  \\
        \hline
       Stellar Mass ($M_{\odot}$) & \multicolumn{2}{l}{{0.871}} & & \\
    Stellar Radius ($R_{\odot}$) & \multicolumn{2}{l}{{0.852}} & & \\

       \\
     \hline
       {Planet Parameters}  &  .04 & b & c & d \\
        \hline
       Period (days) & 0.528474 & 4.65382 & 9.15067 & 19.9807 \\
       Semi-Major Axis (AU) & 0.01222 & 0.05210 & 0.0818 & 0.1376 \\
       Mass ($M_{\bigoplus}$) & 2.65 & 8.5 & 8.6 & 9.5 \\
       Inclination (degrees) & 72.80 & 88.99 & 88.52 & 88.753 \\
       Eccentricity & 0$^{1}$ & 0.183 & 0.065 &  0.075 \\
   Time of Conjunction (BJD) & 2458350.8394 & 2458355.35520 & 2458352.7582 & 2458342.8514 \\
    \end{tabular}
   
    \caption{Table containing all necessary planetary and stellar parameters to perform both secular theory and n-body integrations. All values reported come from \citep{Quinn2019} and are the median values drawn from the ExoFastv2 fit of that paper. $^{1}$: The eccentricity of the inner planet was set to 0 in the ExoFastv2 fit in \citet{Quinn2019}, and we use that value here as well. }
    \label{tab:ParamsTable}
\end{table*}

\end{document}